\documentclass[pra,preprint,showpacs]{revtex4}
\usepackage[centertags]{amsmath}
\usepackage{amsfonts}
\usepackage{amssymb}
\usepackage{amsthm} 
\usepackage{newlfont}
\usepackage{epsfig}
\usepackage{amscd}
\usepackage{graphicx}
\usepackage{epsfig}
\usepackage{color}
\usepackage{amscd}

\newcommand{\beq}{\begin{equation}}
\newcommand{\eeq}{\end{equation}}
\newcommand{\ba}{\begin{array}}
\newcommand{\ea}{\end{array}}
\newcommand{\bea}{\begin{eqnarray}}
\newcommand{\eea}{\end{eqnarray}}
\begin{document}

\begin{center}
{\large \sc \bf { Remote one-qubit state   control by pure initial state of
two-qubit sender. 
Selective-region- and eigenvalue-creation.
}}

\vskip 15pt

{\large 
G.A. Bochkin and A.I.~Zenchuk 
}

\vskip 8pt

{\it Institute of Problems of Chemical Physics, RAS,
Chernogolovka, Moscow reg., 142432, Russia},\\

\end{center}


\begin{abstract}
We study the problem of remote one-qubit mixed  state creation using a pure initial  state of  
two-qubit sender and spin-1/2 chain as a connecting line. We  express 
the parameters of creatable states in terms of 
 transition amplitudes.  
We show that the creation of complete receiver's state-space 
can be achieved  only
in the chain engineered for the  one-qubit   perfect state transfer (PST) (for instance, in the 
fully engineered Ekert chain), the chain can be arbitrarily long in this case. 
As for the homogeneous chain, the creatable receiver's state region decreases quickly
with the chain length. Both 
homogeneous chains and chains engineered for  PST can be used for the purpose of selective state creation,
when only the restricted part of the whole receiver's state space is of interest.
 Among the parameters of  the receiver's state, the eigenvalue is the most  
hard creatable one and therefore deserves the special study.   
Regarding the homogeneous spin chain, an arbitrary eigenvalue can be created 
only if the chain is of no more than 34 nodes.
Alternating chain allows us to increase this length up to 68 nodes.
 \end{abstract}

\maketitle

\section{Introduction}
\label{Sec:Introduction}

The problem of remote creation of a particular quantum state is one of  fundamental problems in
quantum communication. 
Its prototype is the
pure  quntum state transfer problem,
which  was first formulated
in  well-known paper by Bose \cite{Bose}  for the homogeneous ferromagnet spin chain 
with isotropic  Heisenberg interaction.  Now the state transfer  represents a special direction in
quantum information processing.
Among the spin systems engineered for the   either perfect  or high-fidelity (probability)
one-qubit pure 
 state transfer, we mention such   well-known systems  as the  spin chains
with properly adjusted 
coupling constants  (or the  fully engineered spin chains) \cite{CDEL,ACDE,KS}
and  the  homogeneous  chains with remote end nodes
(the boundary-controlled \cite{GKMT,WLKGGB} and 
optimized boundary-controlled  \cite{NJ,SAOZ} spin chains).  In addition, the 
experimental realization of the perfect state transfer through the three-qubit chain in trichloroethylene is proposed in
\cite{ZLZDLL}.

Studing the perfect state transfer (PST)  problem in spin chains shows its  sensitivity 
to the chain parameters. Moreover, been achieved for a  model system 
(such as the nearest neighbor XY Hamiltonian in \cite{CDEL,ACDE,KS}), it becomes destroyed by 
imperfections, such as remote node interactions and quantum noise, which always reduce the state transfer 
fidelity \cite{CRMF,ZASO,ZASO2,ZASO3,SAOZ} so that the original state can not be perfectly  
transfered between the ends of a chain. As a consequence, the high-fidelity/probability state 
transfer becomes more popular in comparison with the PST,
which is justified in  numerous papers concerning different aspects of  this subject, such as the entanglement 
\cite{Wootters,HW,P,AFOV,HHHH} 
transfer through a quantum chain \cite{DFZ,DZ,BACVV2010,BACVV2011,LS}, 
the entanglement creation  between distant qubits  \cite{BBVB,BZ}, 
the so-called ballistic quantum state transfer \cite{Banchi}, the
high-dimensional state  transfer \cite{YGQ,YGQ2}, the
robustness of state transfer \cite{CRMF,ZASO,ZASO2,ZASO3,QWZ}.

Nevertheless, the   search for  alternative ways of quantum communications  free of the destructive
 effect of imperfections becomes more and more attractive. 
Thus, the so-called information transfer 
was proposed in  \cite{Z_2012}. In this case
we take care of  transfer of all  the  state's parameters 
(instead of the quantum state itself) from the  sender to the receiver. 
These parameters appear linearly in the receiver's state, so that 
we have to solve a system of 
linear algebraic equations to obtain these parameters on the receiver's side.
In turn, this requires  non-quantum mechanical tool, which is a price for the 
robustnesses of the information transfer.   The  conclusion about robustness 
 is based on a simple observation that 
any imperfection of the model changes the coefficients in the above linear system 
without changing the transferred parameters  
(as for the noise, also the averaged effect leads to such change of coefficients).
Consequently, unlike the state transfer, 
this process  is not sensitive to the parameters of the spin chain as well as to the imperfections of the 
experimental realization of the proposed model (nearest neighbor XY Hamiltonian was used in \cite{Z_2012}). 

In  recent paper \cite{Z_2014}, the principles of  
  both perfect state transfer 
\cite{Bose,CDEL,ACDE,KS,KF,KZ_2008,GKMT,FZ,WLKGGB}
and  state-information transfer \cite{Z_2012} were realized in the mixed  
state creation algorithm using short homogeneous 
spin-1/2 chains with nearest neighbor XY-interactions. The basic idea of that paper is 
to handle the parameters of the  creatable state of the remote subsystem (receiver) 
varying the parameters of another subsystem (sender) through the local unitary transformations of the latter. 
Notice, that the most of earlier experiments 
realizing the remote state creation use photons as   carriers of  quantum information 
\cite{ZZHE,BPMEWZ,BBMHP,PBGWK2,PBGWK,DLMRKBPVZBW,XLYG}. State creation  based on spin-1/2 chains 
is suitable 
for the state/information transfer/creation over the relatively short distances, 
for instance, inside of a particular 
quantum device.

 In this paper we represent the detailed study of the
 remote one-qubit mixed state creation of the receiver (the last node of a chain) 
 through the long  spin-1/2 chain
  with  a pure initial state of (at most)
 two-qubit sender (the first and the second nodes of a chain). 
We call the parameters  varying with the purpose to create the needed receiver's 
state as the control parameters, while 
the parameters of the receiver's state are referred to as 
the creatable parameters. 
In the case of  one-qubit sender, the time is required as one of the 
control parameters needed to create a large region of the receiver state-space. 
Therefore, this case can not be considered as the completely local    control 
(i.e., the control through the local parameters  of the sender's initial state) of   state-creation,
because the required  time instant must be reported to the receiver's side 
(perhaps, through a classical communication channel). 
We concentrate on the completely local control   achievable using the two-qubit 
sender (similar to ref.\cite{Z_2014}). In this case the large creatable region can be covered at a properly fixed 
time instant just varying the parameters of the sender's initial state. In other words, the time 
is not included in the list of control parameters.
For a pure initial sender's state, we 
 express  the parameters of creatable state in terms of 
the transition amplitudes and control parameters, so that the receiver's density matrix acquires the very simple form.  
 We show that the creation of the  complete one-qubit receiver's state-space 
can be achieved  
in the chain engineered for the one-qubit PST  \cite{CDEL,ACDE,KS}.
The chain can be arbitrarily long in this case. 
As for the homogeneous chains, on the contrary,  the creatable region decreases very quickly 
with an increase in the chain length.
Apparently, this is an essential  disadvantage of  homogeneous chains. However we show that 
this disadvantage leads to some
 privileges of such chains   in application to  the problem of 
selective-region state creation, 
when we intend to work with a particular subregion of the receiver's state space. 
Namely, the homogeneous 
chains reduce (or even completely remove) the possibility of a ''parasitic'' state creation outside of 
the required 
subregion, if only this subregion is properly selected. 
Such selective state creation can be considered as  the first step in 
construction of the ''branched'' communication systems having several senders and one common receiver.
Notice, that the chains engineered for the PST do not possess this property, 
 although they can be used in the selective-region state creation as well.

 Following ref.\cite{Z_2014}, by the  state of the  receiver $B$ we mean the reduced density 
 matrix (the marginal matrix)
 \begin{eqnarray}\label{margB}
 \rho^B= Tr_{rest} \rho,
 \end{eqnarray}
 where $\rho$ is the density matrix of the whole system and the trace is calculated over 
 the all nodes except for the last one (receiver's node). 
The one-qubit  receiver's state space is parametrized by three parameters. 
One of them is the  eigenvalue and two others are associated with
the  eigenvectors. In turn, among two latter parameters, there is the phase which has no restriction for 
 creation \cite{Z_2014}, i.e.,
any required value of this phase can be created by a proper choice of the 
control parameters. Another eigenvector parameter can, in principle, be tuned to the required value 
by the local unitary 
transformation of the receiver.
On the contrary, the eigenvalue represents the most hard creatable parameter because it is effected by the evolution 
and it is not sensitive to the 
local unitary transformations of the receiver. Thus the eigenvalue creation deserves a special attention. 
It turns out that any eigenvalue can be created in the homogeneous chain of no more than $N_c=34$ nodes. 
Of cause, any value of this parameter can be created in the fully engineered chains of arbitrary  length, 
but such chains themselves are hard for the practical realization. 
Alternatively, we consider the alternating chain with even number of nodes 
for the purpose of remote eigenvalue creation  and show 
that this chain increases the  parameter $N_c$ up to 68 nodes.

Finally we shall mention that the teleportation 
\cite{BBCJPW,BPMEWZ,BBMHP} can be considered as a prototype of the remote  state creation. 
Teleportation  is aimed on the  long distance transfer of an unknown state using the entangled pairs of bits 
\cite{YS1,YS2,ZZHE}.  An inherent  feature of teleportation is that it requires the classical
channel of information transfer 
as a necessary component of the teleportation protocol, unlike the state transfer/creation problem. 
Set of modifications of the state teleportation/creation protocol can be found in 
\cite{BDSSBW,BHLSW,XLYG,G}.

The structure of this  paper is following. In Sec.\ref{Section:analyt}, we represent the basic 
analytical results concerning the map between the control parameters of  sender's initial state 
and the  creatable parameters of 
the receiver's state. We show, in particular, that the creatable region 
covers the whole receiver's state-space at the time instant of PST. 
The state creation in chains governed by the  nearest neighbor XY Hamiltonian is studied in more details in the same 
section. In Sec.\ref{Section:select}, we study the
 state creation inside of the selected  subregions of the receiver's state space 
 on the basis of homogeneous  and fully engineered Ekert chains. Sec.\ref{Section:lam} is 
 devoted to the eigenvalue creation in the
long homogeneous and alternating chains, as well as in the chains engineered for the PST.
The basic results of paper are collected in 
Sec.\ref{Section:conclusions}.

\section{One-qubit receiver's state creation through pure two-qubit initial sender's state}
\label{Section:analyt}

\subsection{One-excitation evolution}
\label{Section:XY}
In this section we introduce  two requirements simplifying the spin dynamics calculation. 
\begin{enumerate}
\item
 The Hamiltonian $H$ commutes  with  the  $z$-projection of the total spin momentum:
\begin{eqnarray}\label{com}
[H,I_z]=0.
\end{eqnarray}
\item
The initial state is  a superposition of the pure states 
with up to single excited spin.
\end{enumerate}

These two requirements
allow us to consider the dynamics in the following basis of $N+1$ independent vectors 
(instead of $2^N$ independent vectors in general case of  $N$-qubit dynamics):
\begin{eqnarray}
|0\rangle \equiv |\underbrace{0\dots 0}_{N}\rangle,\;\;\;
|n\rangle \equiv |\underbrace{0\dots 0}_{n-1}1\underbrace{0\dots 0}_{N-n}\rangle, \;\;n=1,\dots,N.
\end{eqnarray}
Next, we can write the Hamiltonian in 
the following two-block diagonal form
\begin{eqnarray}
H={\mbox{diag}}(H_0,H_1),
\end{eqnarray}
where $H_0$, is written in the basis of a single vector $|0\rangle$ (thus, it is a scalar) and the block $H_1$ is 
written in the basis $|n\rangle$, $n=1,\dots,N$.  Without the loss of generality, we take $H_0=0$.

Let us consider a pure  initial state $|\Psi_0\rangle$ of our quantum system. 
In accordance with the Schr\"odinger equation,  evolution of this state reads:
\begin{eqnarray}\label{ev0}
|\Psi(t)\rangle = e^{-i H t} | \Psi_0\rangle.
\end{eqnarray}
As known, the receiver's state   is mixed in general and  can be written as (in the case of one-excitation)
\begin{eqnarray}\label{rhoB}
\rho^B \equiv {\mbox{Tr}}_{1,2,\dots,N-1} \rho = \left(
\begin{array}{cc}
1-|f_N|^2 &f_N^* f_0  \cr
f_N f_0^*  & |f_N|^2
\end{array}
\right)= \left(
\begin{array}{cc}
1-R_N^2 &R_N R_0 e^{-2 \pi i(\Phi_N-\Phi_0)} \cr
R_N R_0 e^{2 \pi i(\Phi_N-\Phi_0)} & R_N^2
\end{array}
\right).
\end{eqnarray}
Here the trace is taken over the nodes $1,\dots,N-1$,
star means the complex conjugate value and $f_N$, $f_0$ are the  probability 
amplitudes,
\begin{eqnarray}
f_i&=&\langle i| e^{-i H t} |\Psi_0\rangle = R_{i} e^{2 \pi i \Phi_i},\;\;i=0,\dots,N,
\end{eqnarray}
where $R_i$ and $\Phi_i$ are real parameters and $R_i$ are positive.
Remember a  natural constraint  
 \begin{eqnarray}
\label{constr}
|f_N|^2 +|f_0|^2 \le 1 \;\; \Rightarrow \;\; R_N^2 +R_0^2 \le 1,
\end{eqnarray}
where the equality corresponds to the pure state creation  because in this case  $f_i \equiv 0 $ ($i\neq 0,N$)
and the only nonzero eigenvalue equals one (the last statement   can be directly verified using eq.(\ref{lam}) derived  below). 
This phenomenon is equivalent to the PST  in the case of one-qubit sender.
Constraint (\ref{constr})  suggests the following parametrization of $R_N$:
\begin{eqnarray}\label{RN}
R_N= \sqrt{1-R_0^2} R, 
\end{eqnarray}
therewith
\begin{eqnarray}\label{qube}\label{intR0}
&&
0\le R_0 \le 1,\;\;
\\\label{intR}
&&
0\le R \le 1,\\\label{intPhi}
&&
0\le \Phi \le 1,\;\;\Phi= \Phi_N-\Phi_0.
\end{eqnarray}
 Thus, three parameters $R_0$, $R$ and $\Phi$ (which are defined by the 
 initial state of our quantum system and by the interaction Hamiltonian)
 completely characterize the possible creatable  receiver's state. However, 
 representation (\ref{rhoB}) of the density matrix $\rho^B$
 is not a preferred  one  because it does not give us a simple way to  estimate whether the 
 whole state space of the receiver is creatable. The following factorized  representation
 allows us to realize this estimation  
 giving  the convenient parametrization
 of the receiver's state-space:
\begin{eqnarray}\label{rhoULU}
\rho^B= U^B \Lambda^B (U^B)^+,
\end{eqnarray}
where 
$\Lambda^B$ is the diagonal matrix of eigenvalues and $U^B$ is the matrix of eigenvectors,
which read as follows in our case: 
\begin{eqnarray}
\label{Lambda}
&&
\Lambda^B={\mbox{diag}}(\lambda,1-\lambda),\\\label{U}
&&
U^B=
\left(\begin{array}{cc}
\cos \frac{\beta_1 \pi}{2} & -e^{-2 i \beta_2\pi} \sin \frac{\beta_1\pi}{2} \cr
e^{2 i \beta_2\pi} \sin \frac{\beta_1\pi}{2}  & \cos \frac{\beta_1\pi}{2}.
\end{array}\right)
\end{eqnarray}
with $\lambda$ and $\beta_i$ ($i=1,2$) varying  inside of the intervals 
\begin{eqnarray}
\label{lamint}
&&
\frac{1}{2} \le \lambda \le 1, \\\label{betint}
&&
0\le \beta_i\le 1,\;\;i=1,2.
\end{eqnarray}
Intervals (\ref{lamint}) and (\ref{betint}) cover the whole state-space of the receiver.
Note that the maximally mixed state is characterized by a single parameter $\lambda=\frac{1}{2}$.

 Another advantage of representation (\ref{rhoULU}) is that 
 it  separates the whole parameter space of  receiver's state 
into two parts:
\begin{eqnarray}\label{threeparts}
&&
{\mbox{The independent  eigenvalues of $\rho^B$: the only parameter $\lambda$}},\\\nonumber
&&
{\mbox{The independent eigenvector-parameters of $\rho^B$:  $\beta_1$ and $\beta_2$ }}.
\end{eqnarray}
Obviously, the parameters $\lambda$ and $\beta_i$, $i=1,2$, 
are related with $R_0$, $R$ and $\Phi$ as follows:
\begin{eqnarray}\label{lam}
\lambda&=&\frac{1}{2}
\left(
1+\sqrt{(1-2 R_N^2)^2 + 4 R_N^2 R_0^2 }
\right),\\
\cos\beta_1 \pi &=&\frac{1-2R_N^2}{\sqrt{(1-2R_N^2)^2 +  4 R_N^2 R_0^2}},\;\;\Rightarrow
\\\label{arccosbet1}\label{cosbet1}
\beta_1 \pi &=&\arccos\frac{1-2R_N^2}{\sqrt{(1-2R_N^2)^2 +  4 R_N^2 R_0^2}},
\\\label{beta2}
\beta_2 &=&\Phi
\end{eqnarray}
with $R_N$ from (\ref{RN}).
Clearly, if the triple  $(R_0,R,\Phi)$ 
can run all  points in cube
(\ref{qube}-\ref{intPhi}),  then the triple $(\lambda, \beta_1,\beta_2)$ takes all 
values inside of the cube 
(\ref{lamint},\ref{betint}), and thus the whole   receiver's state-space  is creatable. 
However, this is possible only in special cases (like Ekert chain in Sec.\ref{Section:map}). 
Usually, only a part of the receiver's state-space can be created (see homogeneous chains in Sec.\ref{Section:map}).  
Formulas (\ref{lam}-\ref{beta2}) represent the  map between the control parameters of the sender 
(embedded in $R_0$, $R$ and $\Phi$) and the creatable parameters $\lambda$, $\beta_1$ and $\beta_2$
of the receiver. Now we specify the 
dependence on the control parameters introducing a particular sender's initial state.

\subsection{Two-node sender with one excitation initial state}
\label{Section:2NSender}

 For the purpose of effective remote control of the one-qubit receiver's state,
 we take the  two node sender with the  pure one-excitation initial state of the following general 
 form:
 \begin{eqnarray}\label{sins}
&&
|\Psi_0^S\rangle= a_0 |00\rangle + a_1 |10\rangle + a_2 |01\rangle,\\\label{sinsconstr}
&&
\sum_{i=0}^2 |a_i|^2 =1,
 \end{eqnarray}
 where $a_i$ ($i=0,1,2$) are the control parameters with  constraint (\ref{sinsconstr}). 
 Since the common phase of a pure state  does not effect the density matrix, 
  we take the real positive $a_0$ without the loss of generality.
   Obviously, the above sender's initial state can be obtained 
  from the ground sender's state  $|00\rangle$ 
   using  the following 
  $SU(3)$-transformation \cite{Z_2014}:
  \begin{eqnarray}\label{SU3}
  U^A=\left(
  \begin{array}{ccc}
  a_0 &\displaystyle  -\frac{a_1^*}{\sqrt{1-|a_2|^2}} &\displaystyle  - \frac{a_0 a_2^*}{\sqrt{1-|a_2|^2}}\cr
  a_1 &\displaystyle  \frac{a_0}{\sqrt{1-|a_2|^2}}&\displaystyle  -\frac{a_1 a_2^*}{\sqrt{1-|a_2|^2}}\cr
  a_2&0&\displaystyle \sqrt{1-|a_2|^2}
  \end{array}
  \right),\;\;\sum_{i=0}^2 |a_i|^2=1,
  \end{eqnarray}
  i.e., $|\Psi_0^S\rangle = U^A |00\rangle$. 
  This is a 5-parameter  transformation (one real parameter $a_0$, 
  two independent  amplitudes and two phases  of $a_1$ and $a_2$) 
  and it represents a particular case   of the general eight-parameter
  $SU(3)$ transformation. We will show that transformation (\ref{SU3}) establishes the 
  maximal possible control of the   one-qubit receiver state in the framework of our model.
 Therewith  the rest of the quantum system is in the ground initial state,
 \begin{eqnarray}\label{rest}
|\Psi_0^{rest}\rangle= |\underbrace{0,\dots,0}_{N-2}\rangle 
.
 \end{eqnarray}
 Thus, the initial state of the whole system reads:
\begin{eqnarray}\label{srins}
|\Psi_0\rangle \equiv|\Psi_0^S\rangle \otimes |\Psi_0^{rest}\rangle=
\sum_{i=0}^2 a_i |i\rangle.
\end{eqnarray}

We see that the control capability of the sender can be described in  two equivalent ways: 
by the  parameters of initial state (see eq.(\ref{sins}))
 and by the parameters of  unitary transformation  (\ref{SU3}) 
of the ground sender's state. 
Since the initial state itself seems to be more physical and more practical 
in comparison with the unitary transformation, hereafter we focus on  formula 
 (\ref{sins}).

Obviously, the control parameters appear linearly in   evolution  (\ref{ev0}) of the  state:
\begin{eqnarray}
|\Psi(t)\rangle &=& e^{-iH t} |\Psi_0\rangle = a_0 e^{-iH_0 t} |0\rangle +
\sum_{j=1}^2 a_j e^{-iH_1 t} |j\rangle=
\\\nonumber
&&
a_0  |0\rangle + \sum_{j=1}^2 a_j e^{-iH_1 t} |j\rangle.
\end{eqnarray}
Consequently, the probability amplitudes appearing in the receiver's state (\ref{rhoB}) are also linear functions 
of the control parameters:
 \begin{eqnarray}\label{NN}
f_N(t)&=&\langle N| e^{-i H t} |\Psi_0\rangle = \sum_{j=1}^2 a_j \langle N| e^{-i H_1 t} |j\rangle
=\sum_{j=1}^2 a_j p_{Nj}(t)
,\\\label{N0}
f_0(t)&=&\langle 0| e^{-i H t} |\Psi_0\rangle = a_0\equiv R_0,
\end{eqnarray}
where
\begin{eqnarray}\label{def_chi}
  p_{kj}(t)=\langle k| e^{-iH_1 t}|j\rangle = r_{kj}(t) e^{2 \pi i \chi_{kj}(t)},\;\;k,j>0,
 \end{eqnarray} 
$r_{kj}$ are positive amplitudes  and $2 \pi \chi_{kj}$ ($0\le \chi_{kj}\le 1 $)  are phases of $p_{kj}$. 
 The meaning of $p_{kj}$ is evident. It is the probability amplitude of the excitation  transition
 from the $j$th to the $k$th spin. Emphasize that these probabilities represent the inherent characteristics of the transmission
 line and do not depend on the control parameters of the sender's initial state.
 
 Thus we see that the parameter $R_0$ is identical to  the  parameter $a_0$ 
 of  the initial state (this is a consequence of  condition (\ref{com})) and does not depend on the particular 
 Hamiltonian. 
Consequently, the only Hamiltonian-dependent parameter
in formulas (\ref{lam},\ref{cosbet1}) is $R$.
Being $H$-dependent, this parameter is not  completely controlled by the sender's initial state.

Let us briefly analyze  the dependence of 
$\lambda$ and $\beta_1$ on $R$. 
Calculating the derivative of $\lambda$ with respect to $R$ we find that 
$\lambda$  has the minimum  at $R_{min}=\frac{1}{\sqrt{2}}$, 
\begin{eqnarray}\label{lammin}
\lambda_{min}= \frac{1}{2} \Big( 1 + R_0 \sqrt{2-R_0^2}\Big),
\end{eqnarray}
and reaches its maximal value $\lambda_{max}=1$ at the boundary points $R=0,1$.
Thus it takes   values in the interval 
\begin{eqnarray}
\frac{1}{2} \Big( 1 + R_0 \sqrt{2-R_0^2}\Big) \le \lambda \le 1,
\end{eqnarray}
provided that $R$  takes   values in its interval (\ref{intR}).

Function  $\cos\beta_1$ is a decreasing functions of $R$
taking its maximal value $1$ at $R=0$ and its minimal value $2 R_0^2-1$ at 
$R=1$. Thus, 
\begin{eqnarray}
2 R_0^2-1\le \cos\beta_1 \le   1,
\end{eqnarray}
provided that $R$ takes  values in its interval (\ref{intR}).
At the point $R_{min}=\frac{1}{\sqrt{2}} $ 
 we have 
 \begin{eqnarray}
 \cos\beta_1|_{R=\frac{1}{\sqrt{2}}}  = \frac{R_0}{\sqrt{2 -R_0^2}}.
 \end{eqnarray}

 \subsection{Analysis of creatable region}
 \label{Section:analysis}
 To proceed further, we introduce
  the following parametrization of the sender's initial state (\ref{sins}) 
  (satisfying constraint (\ref{sinsconstr})):
 \begin{eqnarray}
 \label{aalpha}
 a_0=\sin\frac{\alpha_1 \pi}{2}, \;\;\; a_1= \cos\frac{\alpha_1 \pi}{2}\cos \frac{\alpha_2 \pi}{2} e^{2 i \pi \varphi_1} ,\;\;\;
 a_2=\cos\frac{\alpha_1 \pi}{2}\sin \frac{\alpha_2 \pi}{2} e^{2 i \pi \varphi_2},
 \end{eqnarray}
 therewith 
 \begin{eqnarray}\label{alpint}
0\le \alpha_i  \le 1, \;\;0\le \varphi_i\le 1,\;\;i=1,2.
 \end{eqnarray}
 Then 
 \begin{eqnarray}\label{f02}
 && f_0(t)  = \sin\frac{\alpha_1 \pi}{2}\equiv R_0,\\\label{fN2}
 && f_N(t) = \cos \frac{\alpha_1 \pi}{2}\cos \frac{\alpha_2 \pi}{2} r_{N1}(t)e^{2 \pi i(\varphi_1 + \chi_{N1}(t))}  +  
 \cos\frac{\alpha_1 \pi}{2}\sin \frac{\alpha_2 \pi}{2} r_{N2}(t) e^{2 \pi i(\varphi_2+ \chi_{N2}(t)) }.
 \end{eqnarray}
 The parameter $\alpha_1$ fixes $R_0$ inside of  interval (\ref{intR0})
 (thus $R_0$ does not depend on the time $t$ in our case). 
  The amplitude $R_N$ of $f_N$ reaches its maximal possible value at some time instant $t$ 
 if both terms 
 in eq.(\ref{fN2})  have the same phases  at this  time instant, i.e.
 $\varphi_1$ and $\varphi_2$   satisfy  the
  condition
  \begin{eqnarray}\label{ph}
  \varphi_1 + \chi_{N1}(t) = \varphi_2+ \chi_{N2}(t).
  \end{eqnarray}
  For instance
 \begin{eqnarray}
 \varphi_2=\varphi_1 + \chi_{N1}(t) -\chi_{N2}(t).
 \end{eqnarray}
 Therewith we provide any  needed phase  $\Phi$  (\ref{intPhi})
 at any required
 time instant $t$ ($\Phi_0\equiv 0$ according to the formula (\ref{N0}) for $f_0$):
 \begin{eqnarray}
 \Phi(t) \equiv \Phi_N(t)= \varphi_1 + \chi_{N1}(t).
 \end{eqnarray}
So that  any required  phase $\beta_2$ (\ref{beta2}) of the receiver's state can be created. 
 
 Owing to phase-relation (\ref{ph}), we have for $R$:
 \begin{eqnarray}\label{absrN}
 R(t)  = \cos \frac{\alpha_2 \pi}{2} r_{N1}(t) + \sin \frac{\alpha_2 \pi}{2} r_{N2}(t).
 \end{eqnarray}
 Our purpose is to find such parameters of the Hamiltonian  and such time instant $t_0$
 that  $R$ covers the whole interval (\ref{intR}).
 This is possible only if the chain is engineered for the  PST. 
 In general, the following proposition holds.
 
 {\bf Proposition 1.} Let $R$ take some particular value $R_p$ at a given time instant $t=t_1$. 
 Then $R$ takes  values at least in the interval 
 \begin{eqnarray}\label{intRPr1}
 0\le R\le R_p
 \end{eqnarray}
 during the time interval
 \begin{eqnarray}\label{timeint1}
 0\le t \le t_1.
 \end{eqnarray}

 {\it Proof.}
 This statement follows from the fact that 
 $R(0)=0$ and   $R(t)$ is  a continuous function of $t$.
 $\Box$
 
 Obviously,  $R$  takes values in
 the bigger interval 
 \begin{eqnarray}\label{bigint}
 0\le R\le R_{max}
 \end{eqnarray}
 if the value 
 $R_{max}>R_{p}$  is achievable over the  time interval (\ref{timeint1}).
 In other words, the following  consequence holds.

 {\bf Consequence 1.}
 Let $R$ reach the maximal value $R_{max}$ at some instant $t_0$ inside of the time interval $[0,T]$, $t\in[0,T]$.
 Then all receiver's states creatable during the time interval $[0,T]$ 
 can  be created during the shorter time interval $[0,t_0]$, therewith $R$ takes values in interval (\ref{bigint}).

 Thus, if we consider the time $t$ as  one of the control parameters of the state creation process, then 
 we can cover a large region of the receiver's state space
 even with  $\alpha_2=0$. However, this type of remote state creation is not completely 
 controlled by the local parameters of the sender's initial state 
 since 
 the required time instant (as  a control  parameter)  must be transmitted to the receiver's
 side, which complicates 
 the communication. To avoid this complication, we involve the  variable parameter  
 $\alpha_2$ in the initial state (\ref{sins},\ref{aalpha}). In this case, 
 the large  region of the receiver's state space
 can be created at the properly fixed time instant thus making the remote state creation completely 
 controlled by the local sender's initial state (the local control of state-creation), 
 because the above time instant of state registration can be reported  to the receiver's side in advance.
 For the sake of simplicity, we analyse such control for the case of XY Hamiltonian with 
 the nearest neighbor interactions. However, a similar analysis  can be  elaborated in more complicated case as well.

 \subsection{Evolution governed by nearest-neighbor XY-Hamiltonian.}
 
The XY-Hamiltonian with nearest neighbor interaction reads
 \begin{eqnarray}\label{XY}
H=  \sum_{i=1}^{N-1}D_i (I_{ix} I_{(i+1)x} + I_{iy} I_{(i+1)y})=\sum_{i=1}^{N-1}\frac{D_i}{2} 
(I_{i}^+ I_{i+1}^- + I_{i}^- I_{i+1}^+) ,
\end{eqnarray}
where  $D_i$ are  the coupling constants between the 
nearest neighbors, $I_{j\alpha}$ ($j=1,\dots,N$, $\alpha=x,y,z$) is the 
$j$th spin projection on the $\alpha$-axis, and $I_j^\pm = I_{jx} \pm i I_{jy}$. Hereafter we assume the symmetry 
\begin{eqnarray}
\label{symD}
D_i=D_{N-i},
\end{eqnarray}
except for the alternating chain with odd  $N$ in Sec.(\ref{Section:lamcr}).
Obviously, condition (\ref{com}) is satisfied for this Hamiltonian.
Now we formulate the following  Proposition concerning the local control of the 
remote state creation.   
 
 {\bf Proposition 2.} 
  Let 
  the function $r_{N1}(t)$ take the maximal value 
  $r_{max}$ at $t=t_0$, $r_{N1}(t_0)=r_{max}$, and $\chi_{N1}(t_0) \neq 
  \chi_{(N-1)1}(t_0) +\frac{1}{2} n$, ($n=0,\pm 1$) .
 Then $R$  takes  values in the interval
 \begin{eqnarray}\label{intRPr12}
 0\le R\le r_{max}
 \end{eqnarray}
 when 
 \begin{eqnarray}\label{intalp2}
 0\le \alpha_2 \le 1
 \end{eqnarray}
 at the fixed time instant $t_0$. 
 
 {\it Proof.}  The evolution of the pure quantum state is governed by the  Schr\"odinger equation 
 \begin{eqnarray}\label{Ht}
 i |\Psi\rangle_t = H |\Psi\rangle.
 \end{eqnarray}
 Since we deal with the nearest neighbor Hamiltonian (\ref{XY}),
its nonzero  elements  read:
 \begin{eqnarray}
 H_{i,i+1}=H_{i+1,i}=\frac{D_i}{2},\;\;i=1,\;\;N-1.
 \end{eqnarray}
Let us consider the initial state $|\Psi(0)\rangle=|1\rangle$. Then the 
last row of eq.(\ref{Ht}) can be written in terms of transition amplitudes (\ref{def_chi}):
\begin{eqnarray}\label{Htlast}
 i \frac{d}{d t}p_{N1} = \frac{D_{N-1}}{2} p_{( N-1)1}.
 \end{eqnarray}
 The complex conjugate of this equation reads 
 \begin{eqnarray}\label{Htlastcc}
 i \frac{d}{d t}p^*_{N1} = -\frac{D_{N-1}}{2} p^*_{( N-1)1}.
 \end{eqnarray}
 Multiplying eqs.(\ref{Htlast}) and (\ref{Htlastcc}) by, respectively, 
 $p_{N1}^*$ and $p_{N1}$ and adding  them we obtain 
 \begin{eqnarray}
  \frac{d}{d t}(r_{N1})^2 =  D_{N-1} r_{N1}r_{( N-1)1} \sin(2 \pi (\chi_{(N-1)1}-\chi_{N1})).
 \end{eqnarray}
At the 
time instant $t=t_0$ corresponding to the extremum of $r_{N1}(t)$ 
 we have 
 \begin{eqnarray}\label{pN1}
 &&
 \frac{d}{d t} r_{N1}(t_0)=0, \;\;{\mbox{consequently}} \\\label{pN2}
 &&
  r_{N1}(t_0)r_{( N-1)1}(t_0) \sin\Big(2 \pi (\chi_{(N-1)1}(t_0)-\chi_{N1}(t_0))\Big)  =0.
 \end{eqnarray}
Since $ \chi_{N1}(t_0)\neq \chi_{(N-1)1}(t_0) +\frac{1}{2} n$, ($n=0,\pm 1$) 
by our assumption and $r_{N1}(t_0)>0$, we obtain 
  \begin{eqnarray}\label{pN22}
 &&
   r_{(N-1)1}(t_0) =0.
 \end{eqnarray}
 In view of the Hamiltonian symmetry we can write 
 $r_{ij}=r_{ji} = r_{(N-i+1)(N-j+1)}$, then eq.(\ref{pN22}) yields
 \begin{eqnarray}
 r_{(N-1) 1}(t_0) = r_{N2}(t_0)= 0.
 \end{eqnarray}
 This means that  condition (\ref{ph}) is automatically satisfied  at the extremum point of $r_{N1}$
 (because the right hand side of eq.(\ref{fN2}) has only one term at $t=t_0$).
 Thus, $R(t_0)$ defined by eq.(\ref{absrN}) reads
 \begin{eqnarray}\label{absrNnn}
 R(t_0)  = 
 \cos\frac{\alpha_2\pi}{2} r_{max} 
 \end{eqnarray}
and  takes  values  in 
 interval (\ref{intRPr12}) if $\alpha_2$ varies inside of interval (\ref{intalp2}). 
 $\Box$

 In this case we have the completely local  state creation control with three control  
 parameters $\alpha_1$, $\alpha_2$ and $\varphi_1$. 
We shall give the following   remarks.
 \begin{enumerate}
 \item
 Hereafter we refer to the above time instant $t_0$ 
 as the time instant of highest-probability state transfer (therewith the highest-probability state transfer 
 is not always the high-probability state transfer \cite{KZ_2008} because this probability (i.e. $r_{max}$) 
 can be far from one).
 The creation of the  complete receiver's state space 
 is possible if $r_{max}=1$, i.e., if the chain is engineered for the pure one-qubit  PST.
 \item
 The conditions of Proposition 2 hold for the  homogeneous and Ekert chains
 whose  evolution is governed by   the  $XY$ 
 Hamiltonian (\ref{XY}) with the nearest neighbor interactions  at least over the  time intervals $\sim N$ 
 considered here. 
 In this case  
   two control parameters $\alpha_1$ and $\alpha_2$
  are enough to cover the maximal region in the space $(\lambda,\beta_1)$, therewith the parameter $\varphi_1$ 
  is responsible for $\beta_2$ and has no influence on $\lambda$ and $\beta_1$.
 In other cases 
 the phases $\phi_i$, $i=1,2$, 
 (see eqs.(\ref{aalpha})) must be included  into the set of active control parameters 
 to obtain  similar results.
 \end{enumerate}
 
 In Secs.\ref{Section:map} and \ref{Section:select}, we consider only 
 the two-parameter state creation control, i.e., the creation of 
 $\lambda$ and $\beta_1$ varying   $\alpha_1$ and $\alpha_2$, and
 explore several features of such control. The creation of $\beta_2$ is trivial and will 
 not be considered below.

 \subsection{State creation algorithm as a map of control parameter-space
 $(\alpha_1,\alpha_2)$ into 
 creatable parameter-space $(\lambda,\beta_1)$.} 
 \label{Section:map}
 
\begin{figure*}
   \epsfig{file=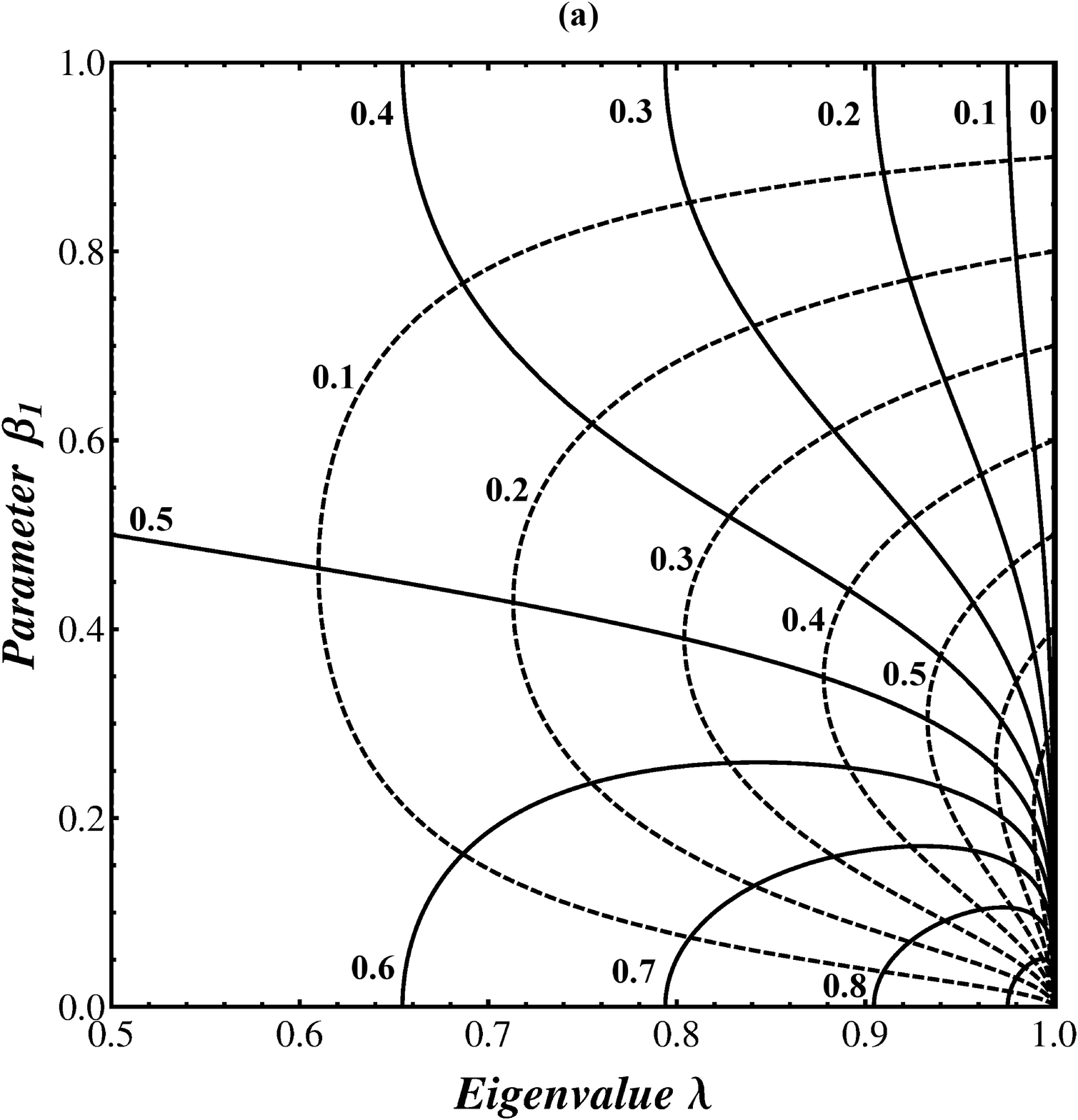,
  scale=0.08
   ,angle=0
}\epsfig{file=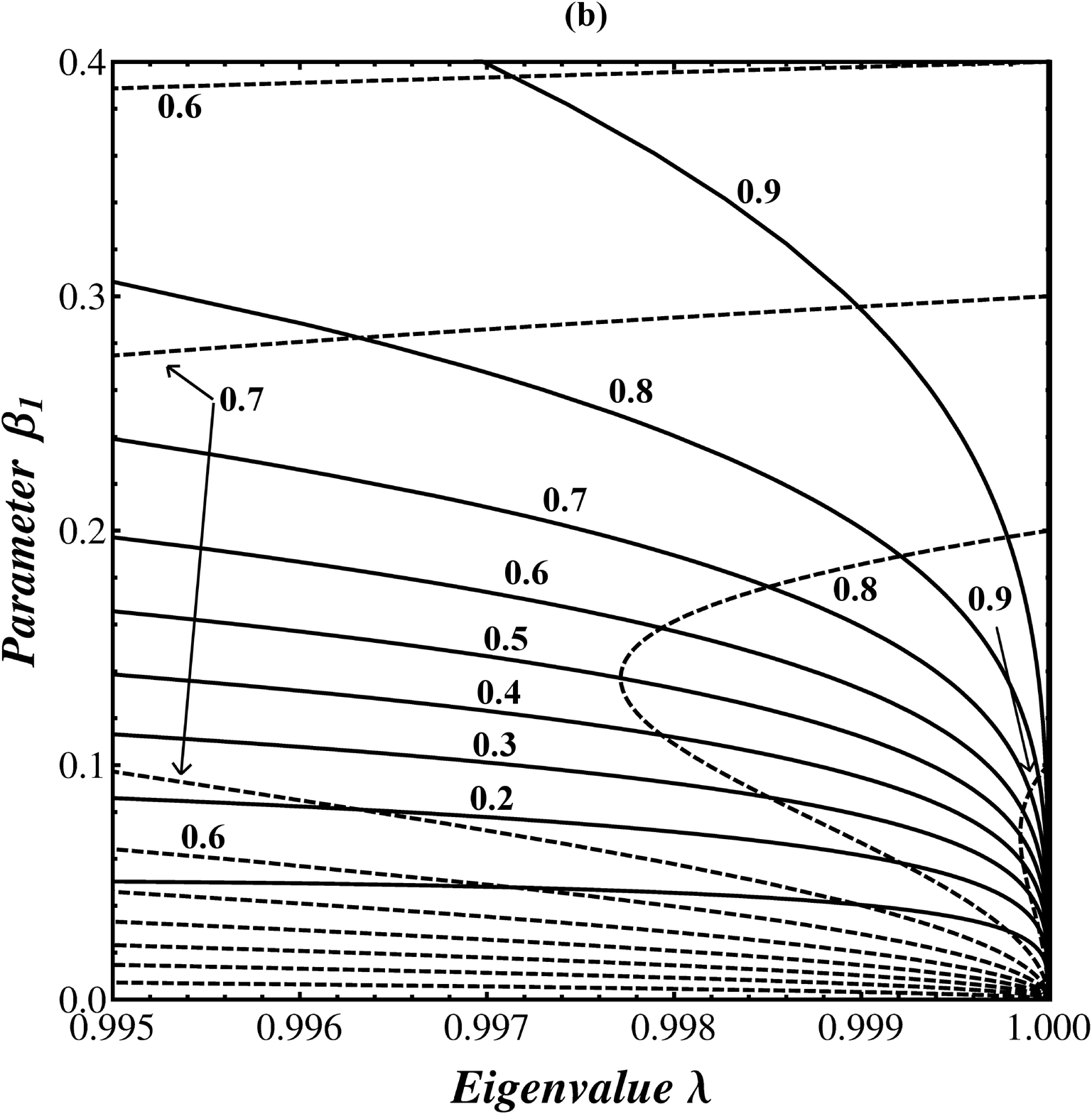,
  scale=0.08
   ,angle=0
}

\caption{The map $(\alpha_1,\alpha_2)$ $\to$  $(\lambda,\beta_1)$ 
in the chains engineered for the pure  one-qubit PST (for instance, in the Ekert chain with $t_0=\pi$). Solid  and dashed 
lines correspond to the  constant values of the parameters, respectively,  $\alpha_2$ and $\alpha_1$,  
appropriate values of these parameters are indicated in the picture. 
The variation intervals (\ref{alpint}) of $\alpha_i$
are splitted into 10 equal segments, i.e., 
the parameter increment between the two neighboring lines is 0.1 in both families of curves. 
The solid line  $\alpha_2=0$ coincides with the right vertical coordinate axis,
the dashed line
$\alpha_1=0$  is disrupted and coincides  with the upper and lower horizontal coordinate axes.  
The  solid  and dashed  lines  with, respectively, 
$\alpha_2=1$ and $\alpha_1=1$  shrink to the point $(\lambda,\beta_1)=(1,0)$. The properly scaled 
neighborhood of this point 
is depictured in Fig. (b).
We keep the same gridding in all pictures below. 
} 
  \label{Fig:map} 
\end{figure*}

\begin{figure*}
   \epsfig{file=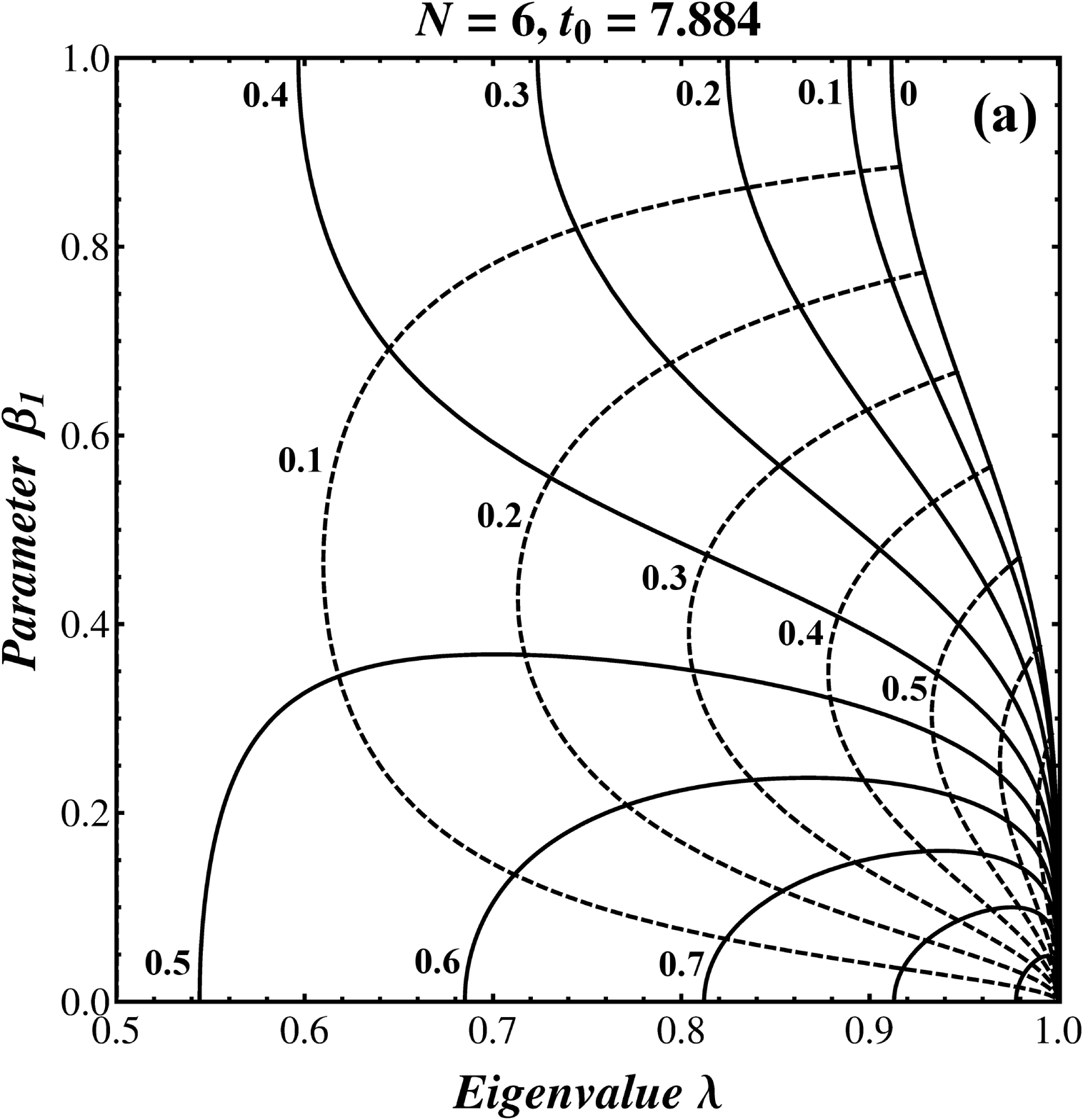,
  scale=0.06
   ,angle=0
}
\epsfig{file=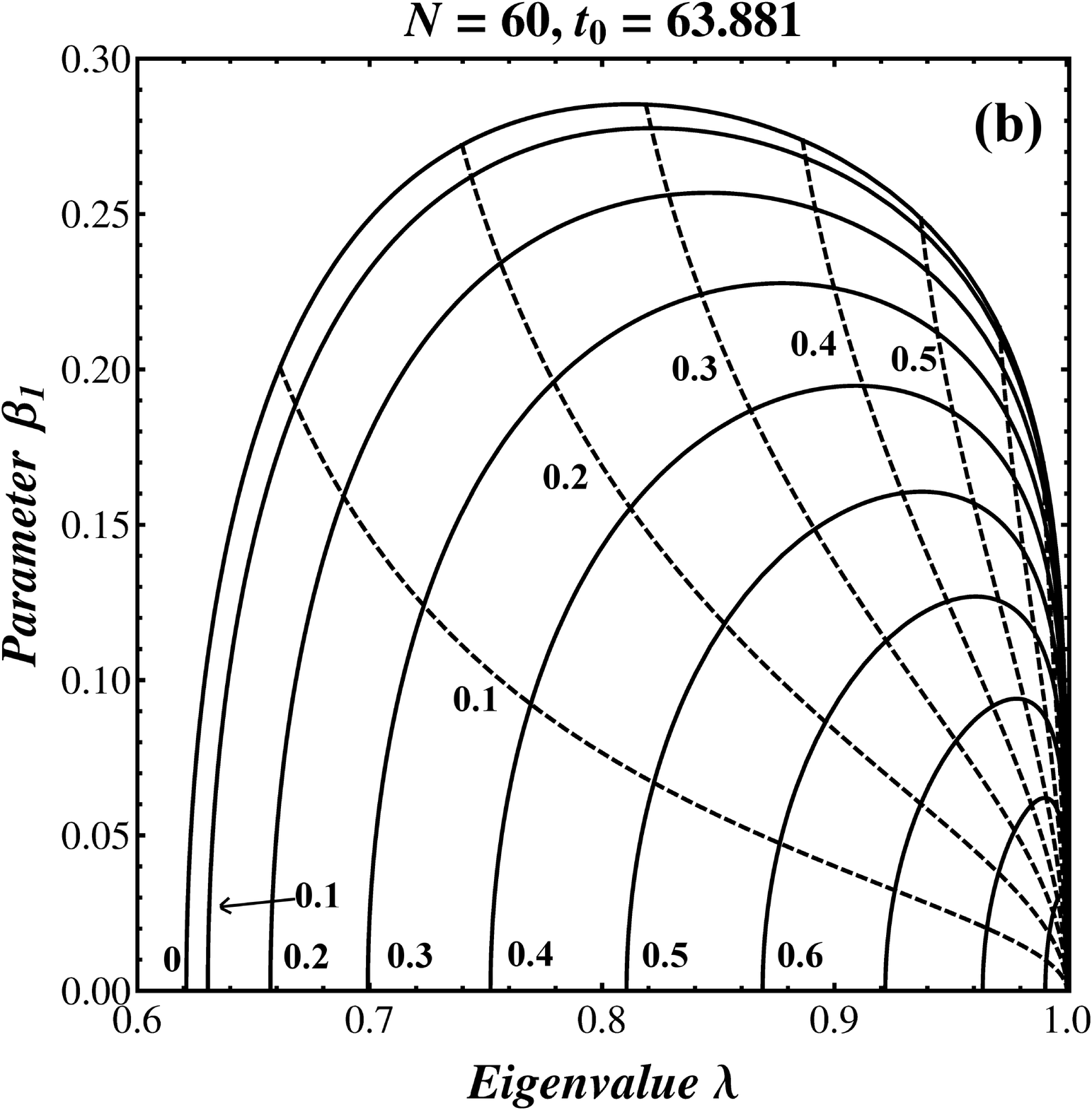,
  scale=0.06
   ,angle=0
}
 \epsfig{file=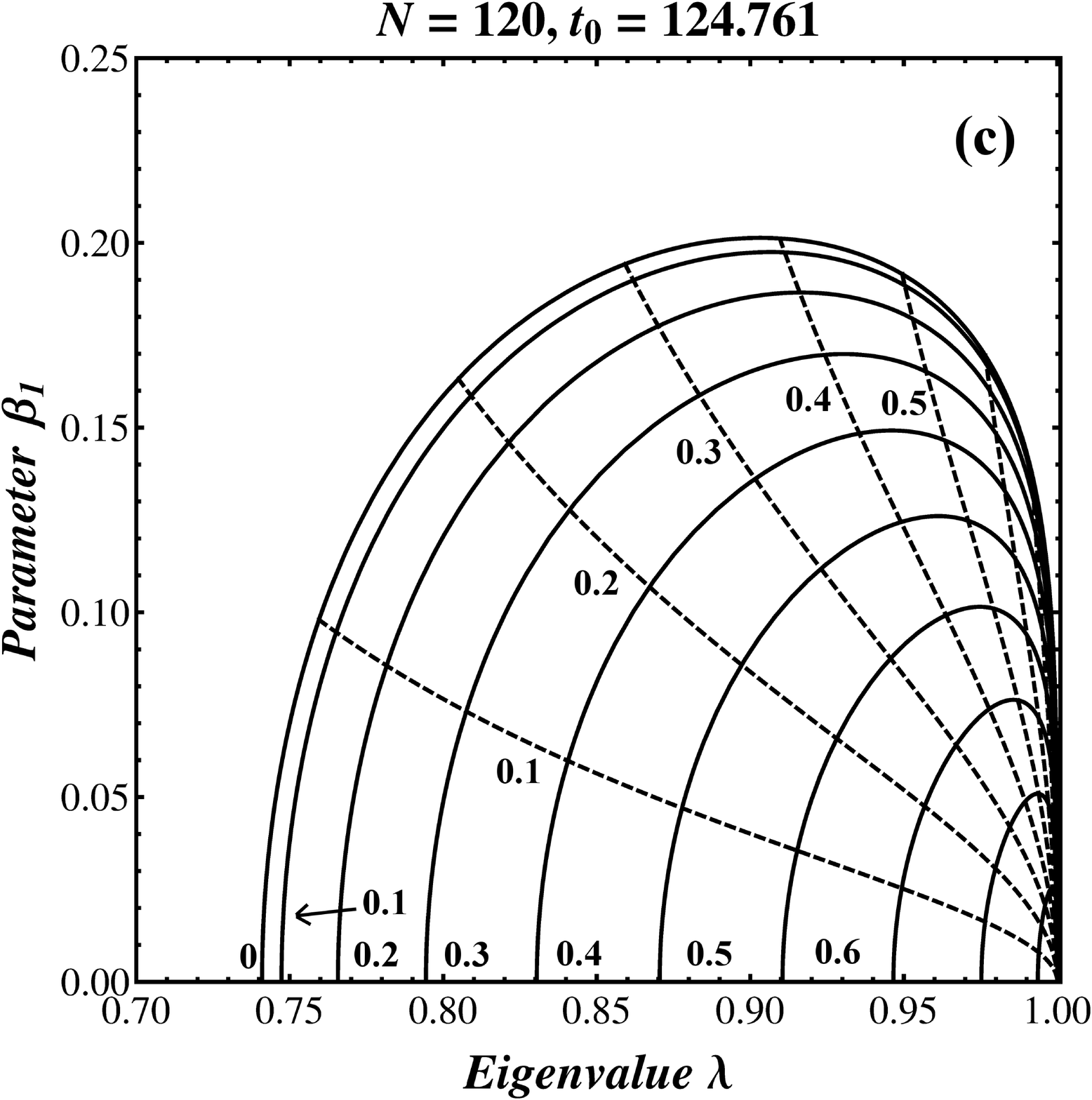,
  scale=0.06
   ,angle=0
}
\caption{The map $(\alpha_1,\alpha_2)$ $\to$  $(\lambda,\beta_1)$ in the homogeneous 
chains with  $N=6$, 60, 120, at the time instants  $t_0$ corresponding to the maximum of $p_{N1}$
(the highest-probability state transfer). 
} 
  \label{Fig:hom} 
\end{figure*}

  If we need to create a receiver's state inside of a particular
subregion of the whole receiver's state space, we need to know the appropriate parameters of the 
sender's initial state. 
This suggests us to consider 
the state creation algorithm  as a map of parameters of arbitrary sender's initial state 
(the so-called control parameters, $\alpha_1$ and $\alpha_2$ in our case) to the set of  parameters of  the receiver's state space 
(the so-called creatable parameters, $\lambda$ and $\beta_1$ in our case).
This map  is depicted in Fig.\ref{Fig:map} 
for the ideal case of the completely creatable receiver's space.
As mentioned above, this situation can be realized in the 
chains engineered for the 
PST.  The Ekert chain can be considered as an example \cite{CDEL}, then
\begin{eqnarray}
D_i=\sqrt{i (N-i)}
\end{eqnarray}
in eq.(\ref{XY}), therewith 
$t_0=\pi$.
In the case of homogeneous spin chain ($D_i \equiv D$, $i=1,\dots,N-1$, in eq.(\ref{XY});
we put $D=1$ for simplicity (dimensionless time $t$)), 
the map differs from that shown in Fig.\ref{Fig:map}.
We observe that the creatable region  of the receiver's  space 
decreases very quickly with the chain length, as
shown in Fig.\ref{Fig:hom}, where chains of 
6, 60 and 120 nodes are considered at the time instants, respectively, 
$t_0= 7.884$, 63.881 and 124.761.
Figs.\ref{Fig:map} and \ref{Fig:hom} help us to select the needed region 
in the  control parameter space  $(\alpha_1,\alpha_2)$  to create the state inside of the  required region of the 
creatable parameter-space  $(\lambda,\beta_1)$.  

\section{Selection of  creatable  subregions.}
\label{Section:select}
We have considered a problem of maximal possible covering of the receiver's state space. It is shown 
 for  Hamiltonian (\ref{XY}) that 
the maximal creatable region corresponds to the time instant $t_0$ of 
the highest-probability state transfer. 
Any deviation from $t_0$ reduces  the creatable region. 
However, this unpleased phenomenon turns out to
be useful if we would like to work only with a restricted subregion of the  receiver's
state space without interacting with its 
remaining part. 
For instance, this problem appears in a ''branched'' communication line when we need to share the creatable
region among several 
senders, so that each sender works only with its own subregion. In fact, Fig.\ref{Fig:hom} shows,  that 
the creatable region of  homogeneous chain of 120 nodes is restricted, roughly speaking, by the rectangle 
$0.74<\lambda\le 1$, $0\le \beta_1 < 0.21$.  
So, the region outside of this rectangle  can be safely used for other purposes.

 Now we describe the separation of several non-overlapping creatable  subregions. 
 Our results  are 
based on the following observation.
 If we take $t_1<t_0$, then the conditions of Proposition 2 are broken, so that we do not cover
 the maximal creatable region varying the control parameters $\alpha_1$ and 
 $\alpha_2$.
 Moreover, 
 the parameter $R$ in formulas (\ref{lam}) 
and (\ref{cosbet1})
can not take all  values  in  interval (\ref{intRPr1}) (remember that $t$ is fixed here, unlike Proposition 1). 
In general,   
the lower and upper  boundaries  appear:
\begin{eqnarray}
q_{min}(t_1) \le R \le q_{max}(t_1). 
\end{eqnarray}
Thus, considering  $K$ chains of different lengths 
$N_i$, $N_1<\dots <N_K$, and appropriate time instant $t_i$
 such that 
\begin{eqnarray}
q^{i}_{min}(t_i) > q^{i+1}_{max}(t_{i+1}), \;\;i=1,\dots, K-1, 
\end{eqnarray}
we may select $K$  non-overlapping creatable subregions in the receiver's state space.
All these regions have the only common point $(\lambda,\beta_1)=(1,0)$. 

\begin{figure*}
   \epsfig{file=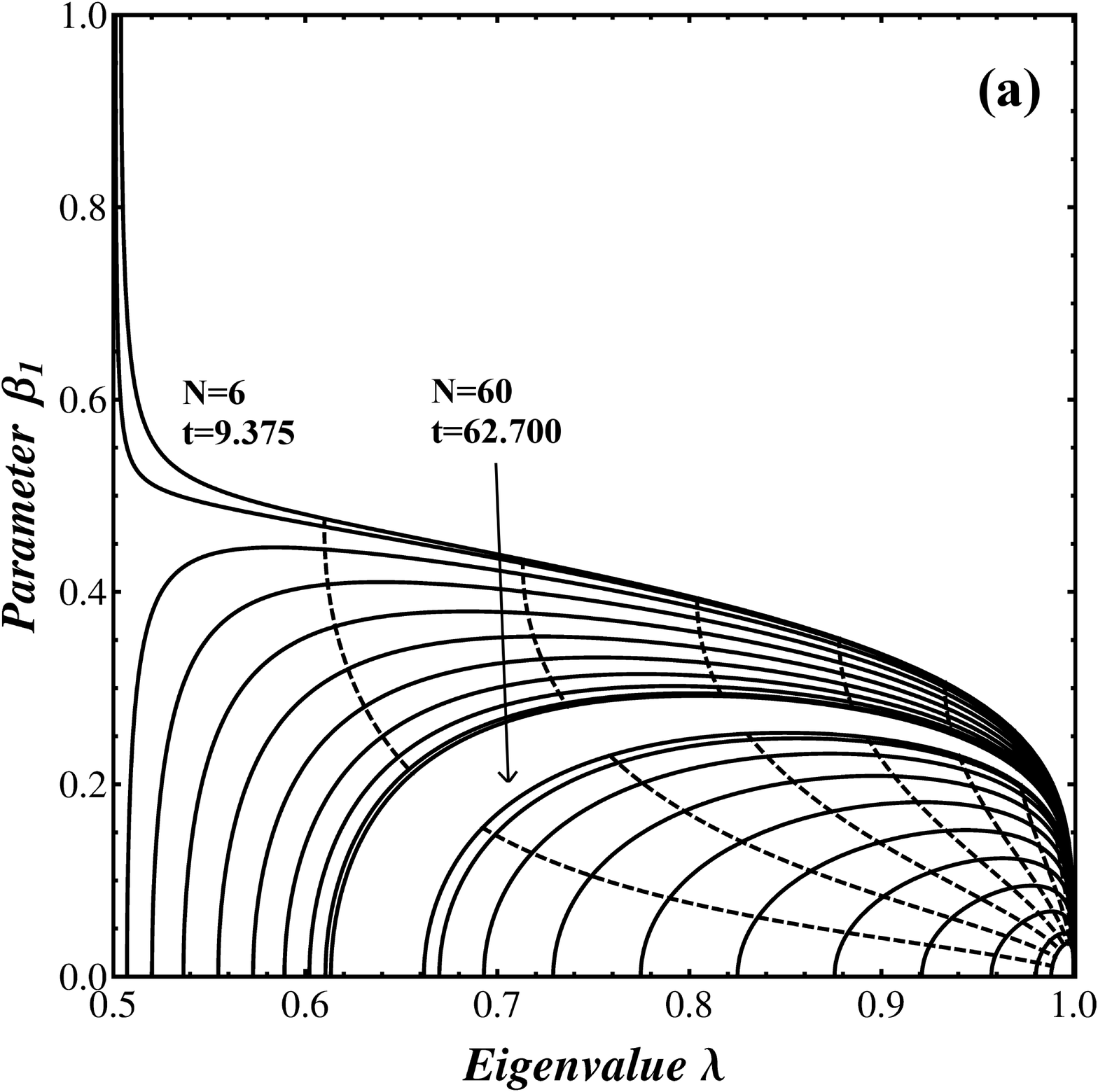,
  scale=0.08
   ,angle=0
}
\epsfig{file=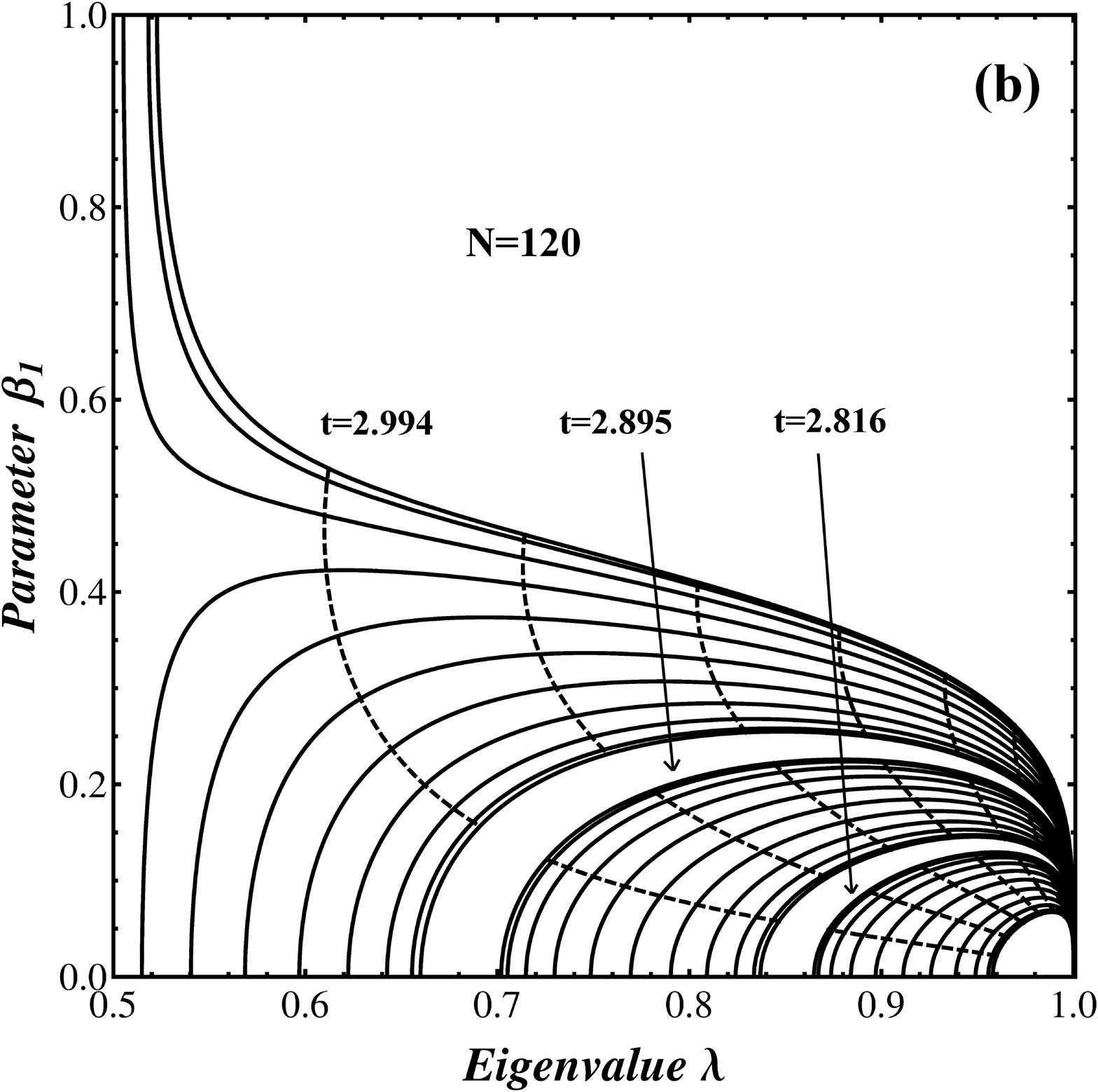,
  scale=0.08
   ,angle=0
}

\caption{The selectivity in the remote state creation. Choosing the proper chain lengths and/or 
the time instants of the state registration 
we obtain different braid-shaped creatable  regions.  Gridding lines inside of each braid
are  the same as  in Fig.\ref{Fig:map}, i.e.,  the parameters $\alpha_i$, $i=1,2$ 
take  values in interval (\ref{alpint}).
(a) The homogeneous spin chains of $N=6$ and 60 with the state-detecting  time instants, respectively, 
$ t=9.375$ and 62.7. (b) The Ekert chain with $N=120$ and the state-detecting time instants 
$t=2.994$, 2.895 and 2.816. 
} 
  \label{Fig:destr} 
\end{figure*}

First, we consider the selective state creation using 
the homogeneous chains. In this case we use the time instants $t_i$  and the chain lengths $N_i$ 
as parameters selecting the proper 
creatable subregions. Combining both these 
parameters we can vary the creatable subregion in a needed way.
Example of two  particularly  selected creatable  subregions  corresponding 
to $(N,t)=(6,9.375)$ and $(60,62.7)$
 are shown in Fig. \ref{Fig:destr}a. 

Next, we perform the above selection using 
 the  Ekert chain \cite{CDEL}. 
In this case we can create different subregions using the chains of the same prescribed  length $N$
and varying the time instants $t_i$  of the state registration.
Example of three creatable subregions  corresponding to the chain of $N=120$ spins
and the registration time instants  $t=2.994$, 2.895 and 2.816  are shown in Fig. \ref{Fig:destr}b.

The privilege of homogeneous chains is that their creatable regions are
restricted as shown in Fig.\ref{Fig:hom}, 
which reduces the possibility of ''parasitic'' state creation from  an ''alien'' sender.  
For instance, the sender responsible for the lower region in
Fig.\ref{Fig:destr}a ($N=60$) can not create the states in the upper 
selected region (corresponding to $N=6$)
regardless of the  values of the control parameters,
this conclusion follows from comparison of Figs. \ref{Fig:destr}a and \ref{Fig:hom}b. 
Although the opposite is not true and 6-node chain can
create the ''parasitic'' states in the lower subregion of  Fig.\ref{Fig:destr}.

\section{Long distance eigenvalue creation in homogeneous and non-homogeneous chains}
\label{Section:lam}

 \subsection{Three types of creatable parameters.}
 \label{Section:threepar}
In Secs.\ref{Section:analysis}, \ref{Section:map}
we show that   the whole receiver's state-space \cite{Z_2014} can not be remotely created 
using arbitrary spin chain. 
 But are there equivalent  obstacles for creation of each of  three parameters 
$\lambda$, $\beta_1$ and $\beta_2$ of the receiver's state space?
It seamed out that all these  parameters  behave differently in the state creation process. 

First of all, we shall emphasis the principal difference between the 
eigenvalue $\lambda$ and the  eigenvector-parameters 
$\beta_i$, $i=1,2$. The latter have such an advantage that they, in principle, can be tuned to the 
required values by the local 
unitary transformation of the receiver (assuming that unitary  transformations 
are applicable on the receiver side), which is a quantum-mechanics operation. 
 Indeed, if we have created   the  receiver's state in the form (\ref{rhoULU}), i.e.,
$\rho^B=  U^B \Lambda^B ( U^B)^+$,  while the required  state is $\rho^{req}=  \tilde U^B \Lambda^B ( \tilde U^B)^+$
(for the one-qubit receiver, the unitary transformation $\tilde U^B$ has the form (\ref{U}) with different parameters),
then 
$\rho^{req}= \tilde U^B (U^B)^+\rho^B U^B( \tilde U^B)^+$, i.e., 
the mentioned above local transformation of the receiver reads: 
\begin{eqnarray}
U^{loc}=\tilde U^B (U^B)^+.
\end{eqnarray}
Notice that the transformation $U^{loc}$ depends on the sender's control parameters, which are 
included into $U^B$. Consequently, the receiver needs  information about (some of) the control parameters to 
apply the proper $U^{loc}$. This information must be transfered from the sender to the receiver 
using some additional (classical) 
communication channel, similar to the teleportation algorithm.
This means that, involving $U^{loc}$ into the state-creation algorithm, we lose the completely 
remote control of all
parameters of the receiver state, except for the eigenvalues (matrix $\Lambda^B$) which 
can not be changed by $U^{loc}$. In this paper we do not consider
the local transformations of  the receiver  as a part of the state-creation algorithm.

It is also shown in Sec. \ref{Section:analysis} that the most reliable 
parameter is the phase $\beta_2$ \cite{Z_2014}, because 
any its value can be created using the phases 
$\varphi_i$, $i=1,2$, of the sender's
initial state.
Moreover, this property of the parameter $\beta_2$ does not depend on the 
Hamiltonian governing the spin dynamics (this can be simply demonstrated).
All this suggests us to consider this parameter as a preferable candidate 
for the carrier of quantum information.

Thus, the eigenvalue $\lambda$ turned out to be the most defenseless parameter, 
because (i) we are not able to create its arbitrary value (in general) and (ii) 
it can not be changed by the local 
unitary transformations of receiver. Therefore the eigenvalue is completely defined by 
the sender's initial state  and by the interaction Hamiltonian, and consequently the eigenvalue-creation
deserves a special study.

Let us consider the $\lambda$-creation in more details using three types of chains:
the homogeneous chain, the alternating chain  and 
the chain engineered for the one-qubit PST (Ekert chain).


\subsection{Eigenvalue creation using Ekert chain,  homogeneous and  alternating chains }
\label{Section:lamcr}
Considering the state-creation based on a spin chain of general position, 
the maximal variation interval (\ref{lamint}) for $\lambda$ becomes  compressed.
The reason is pointed out in 
 Sec.\ref{Section:analysis}, where the expression for  $\lambda$ as a function of $R$ is represented, 
 see eq.(\ref{lam}). It was shown that 
 the left boundary of   eigenvalue $\lambda_{l}=\frac{1}{2}$  can be created  if  
$R>R_{min}=\frac{1}{\sqrt{2}}$, see eq.(\ref{lammin}). Consequently, this boundary  
is achievable  if only $R_{min}\le r_{max} $ in eq.(\ref{intRPr12}). 
This suggests us to introduce the parameter $\lambda^{cr}_{min}(N)$ indicating the minimal eigenvalue creatable on 
the receiver site as a characteristics of the chain. 

\begin{figure*}
   \epsfig{file=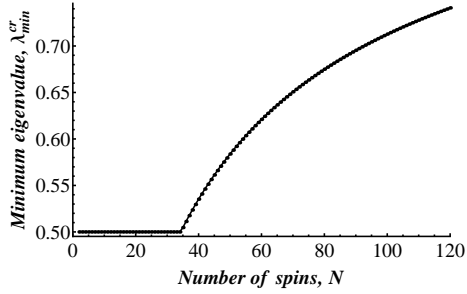,
  scale=0.3
   ,angle=0
}
\caption{The homogeneous spin chains: the minimal creatable eigenvalue $\lambda^{cr}_{min}$ 
as  a function of  chain length 
$N$ .
} 
  \label{Fig:lammin} 
\end{figure*}

Considering the homogeneous spin chain ($D_i =  D_1=1$, $i=1,\dots,N-1$ in eq.(\ref{XY})), we see from  
Fig.\ref{Fig:hom}   that $\lambda_{min}^{cr}(N)=\frac{1}{2}$ only if the chain is short enough
(Fig.\ref{Fig:hom}a), unlike the long chains
(Fig.\ref{Fig:hom}b,c).
The general dependence of $\lambda^{cr}_{\min}$ on $N$ is shown in Fig.\ref{Fig:lammin},
indicating  that  there is such the critical length $N_c^{hom} =34 $ that 
\begin{eqnarray}
\lambda^{cr}_{min}>\frac{1}{2} \;\;{\mbox{for}} \;\; N>N_c^{hom}.
\end{eqnarray}

Of course, $\lambda^{cr}_{min}(N)=\frac{1}{2}$ and does not depend on the chain length in the case of Ekert chain 
($N_c^{PST}=\infty$). 
However, such chains are hard for realization, so that 
we are forced to look for alternative ways of increasing the parameter $N_c$.

\begin{figure*}
   \epsfig{file=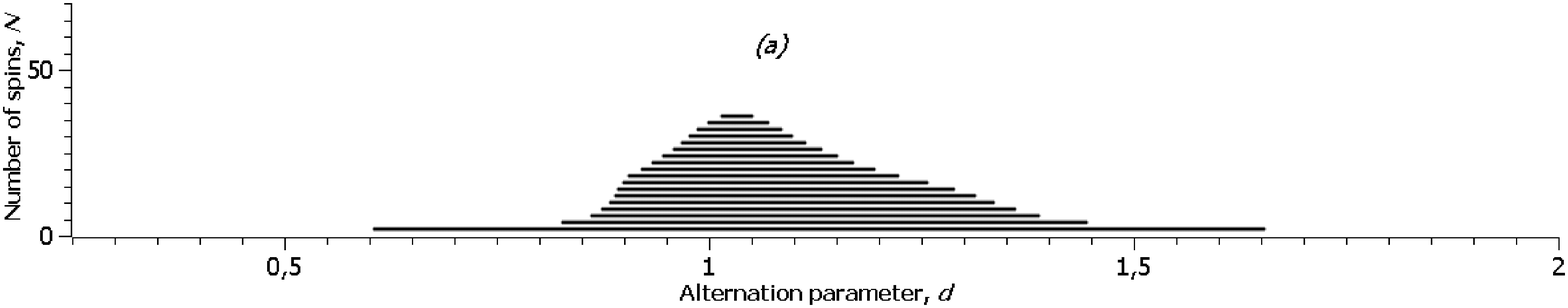,
  scale=0.4
   ,angle=0
}\\
\epsfig{file=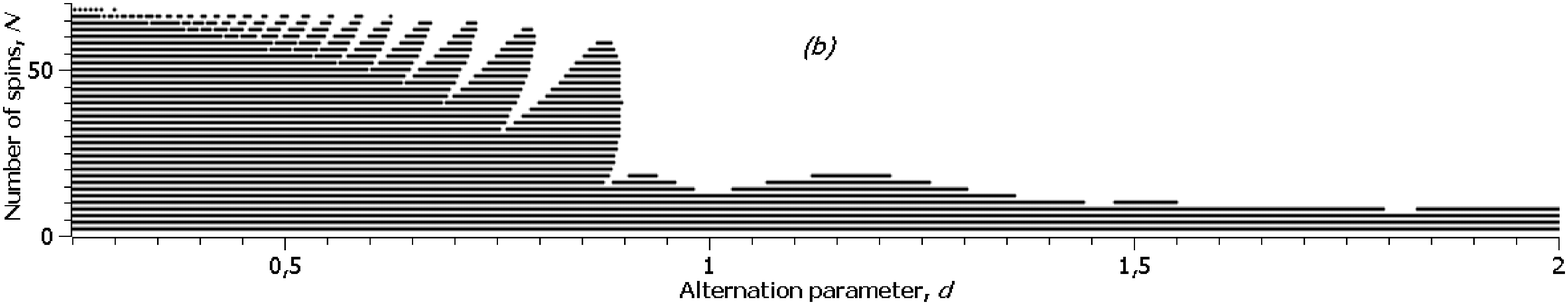,
 scale=0.4
   ,angle=0
}
\caption{The eigenvalue creation in the  alternating spin chains with even $N$ 
over the different  time intervals.
Lines (or spots) correspond to the creation of any $\lambda$ from the interval $\frac{1}{2}\le \lambda \le 1$ 
for the appropriate $N$ and $d$.
The envelope of each figure represents the critical length $N_c^{even}$ as a function of the 
dimerization parameter $d$. 
(a) 
$0\le t\le 1.3 N \min\Big(d,\frac{1}{d}\Big)$, $(N_c^{even})_{max}=36> N_c^{hom} =34$. (b) 
$1.3 \min\Big(d,\frac{1}{d}\Big) <   t\le 1.5 N \max\Big(d,\frac{1}{d}\Big)$, $(N_c^{even})_{max}=68> N_c^{hom}$ .
} 
  \label{Fig:even} 
\end{figure*}

A simple  way to do that 
is using an alternating chain. In this case  $D_i =  D_1=1$, $i=1,3,5,\dots $ and 
$D_i =  D_2=d$, $i=2,4,6,\dots $ in eq.(\ref{XY}). Therewith $d$ is called the 
alternation parameter.
The results of  our calculations for the chain with even number of nodes are collected in 
Fig.\ref{Fig:even}. 
To simplify calculations we put $\alpha_2=0$ in this subsection. Using variable parameter $\alpha_2$ 
we would only slightly modify figures 
without changing the parameter $N_c$.

\begin{figure*}
   \epsfig{file=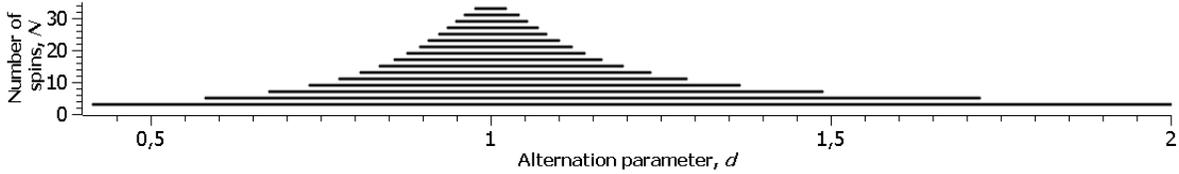,
  scale=0.4
   ,angle=0
}

\caption{The same as Fig.\ref{Fig:even} for the chain with odd $N$
and the time interval 
$0\le  t\le 3 N \min\Big(d,\frac{1}{d}\Big)$; $(N_c^{odd})_{max}= 33<N_c^{hom}$.
} 
  \label{Fig:odd} 
\end{figure*}
The chain with even number of nodes is considered in
 Fig.\ref{Fig:even}. The parameter 
$R$ responsible for the $\lambda$-creation takes its critical value $\frac{1}{\sqrt{2}}$
inside of different time intervals depicted on Figs. \ref{Fig:even}a,b. 
The lines (or spots) mean that any $\lambda$ from the interval 
 $\frac{1}{2}\le \lambda \le 1$ can be created for the proper $N$ and $d$. The envelopes of these figures give 
 the parameter $N_c^{even}$ as a function of $d$. 
 The most reasonable time interval, 
$0\le t\le 1.3 \min\Big(d,\frac{1}{d}\Big)$, is depicted in Figs.\ref{Fig:even}a:
the alternation allows us to increase the length $N_c$ till $N_c^{even}=36$. 
More significant increase in  $N_c$ is observed over the second time interval 
$1.3 \min\Big(d,\frac{1}{d}\Big) <   t\le 1.5 \max\Big(d,\frac{1}{d}\Big)$ shown 
in Figs.\ref{Fig:even}b  (the numerical coefficients $1.3$ and $1.5$ are empiric).
In this case $N_c^{even}=68$ which is twice bigger than $N_c^{hom}$. This  results 
from the   chain ''dimerization''  with decrease in the alternation parameter $d$.
We see that the  function $N_c^{even}(d)$ is not unique 
 for $d<1$ if $R$ achieves its critical value
 $\frac{1}{\sqrt{2}}$ over the time interval corresponding to Fig.\ref{Fig:even}b.

The case of odd $N$ is not interesting because it does not yield any  increase of the critical length 
in comparison with the homogeneous chain
($N_c^{odd} =33 < N_c^{hom}$), as  shown in Fig.\ref{Fig:odd}.

 \section{Conclusions}
 \label{Section:conclusions}
 In this paper we study several aspects of the
  remote state creation using the 
   homogeneous, Ekert and alternating spin-1/2 chains. To simplify calculation, we require the 
   commutation condition (\ref{com}) for the Hamiltonian and one-spin excitation initial state. 
   Based on these requirements are the following results.
   \begin{enumerate}
   \item 
  The receiver's density matrix can be simply expressed in terms of 
  the probability amplitudes. These amplitudes 
 are the  characteristics of the 
  transmission line known in advance, Secs.\ref{Section:XY}, \ref{Section:2NSender}.
  \item
  Three parameters  of the creatable one-qubit state-space can be referred to as 
  the phase and amplitude of the eigenvector  and the eigenvalue. 
  We show that an arbitrary eigenvector's phase can be created using the 
  proper values of the control parameters
  (Sec.\ref{Section:analysis}),
  the eigenvector's  amplitude can be tuned by the unitary transformation of the receiver, while the 
  eigenvalue is most hard creatable and thus deserves the special consideration, Sec.\ref{Section:lam}.
   \item
  Being the most reliably creatable, the eigenvector's phase (parameter $\beta_2$) is a preferable
  candidate for the 
  quantum information carrier in quantum communication lines.
   \end{enumerate}
  In addition, the following results were obtain for  nearest-neighbor XY-Hamiltonian (\ref{XY}). 
   \begin{enumerate}
   \item
   The arbitrary parameters of the two-qubit pure sender's initial state 
   are the control parameters establishing the complete local  control   
    of the receiver's creatable region at the properly fixed time instant of the state registration,
   Secs.\ref{Section:analyt}, \ref{Section:select}.  
  \item
  The maximal creatable region corresponds to the time instant associated with the  
  highest-probability state transfer (although this probability may be far from unit, 
  i.e., ''highest'' does not mean ''high'').
  The creatable region decreases very quickly with the chain length of a homogeneous chain.
  As  anticipated  in \cite{Z_2014},  the complete state space of the one-qubit receiver can be created only 
   in the spin chain engineered for the  pure one-qubit PST, 
   Secs.\ref{Section:analysis}, \ref{Section:map}. 
   \item 
   The map  (control parameter-space) $\to$ (creatable parameter-space) is numerically  described and depicted in 
   Figs.\ref{Fig:map}, \ref{Fig:hom}, thus helping to choose the control parameters needed 
   to create   a particular receiver's state, Sec.\ref{Section:map}. Everywhere  in 
   Figs.\ref{Fig:map}-\ref{Fig:destr} we use the same gridding of the parameter space $(\alpha_1,\alpha_2)$. 
   \item
  Choosing different lengths of the homogeneous chain and different time-instants of the 
  state registration we can select the needed creatable subregion. A similar  result can be achieved using the 
  Ekert chain of a fixed length and different  time-instants of the state registration, Sec.\ref{Section:select}.
  \item
  Considering the process of the remote eigenvalue creation 
  we show that the arbitrary eigenvalue can be created  through the homogeneous spin 
  chain of up to 34 nodes, through the alternating chain of up to 68 
  nodes and through the Ekert chain  of arbitrary length, Sec.\ref{Section:lamcr}. 
   \end{enumerate}
Among the aspects deserving  deeper study we mention (i)
the  transformation of created states
using the tool of local (non-unitary) operations; (ii) the robustness of state-creation with respect to  chaotic permutations and
model imperfections; in particular, the effect of remote spin interactions has to be clarified;  (iii) the model with 
two-excitation initial state (instead of the single-excitation one) has to be explored.

We shall also notice that 
the creation and evolution of quantum correlations is another direction
of quantum information processing stimulating intensive investigations
(for instance, see refs.\cite{DSC,LBAW,NLLZ,ZC,SXSZDWHCKW,RDL}). Currently, the quantum  entanglement 
 \cite{Wootters,HW,AFOV,HHHH} and the quantum  discord \cite{Z0,HV,OZ,Z} 
 are widely accepted measures of quantum correlations. The remote creation of 
entangled quantum  states and states with quantum discord is one more problem 
   postponed for further study.

Authors thank Prof. E.~B.~Fel'dman, Dr. S.~I.~Doronin and Dr. A.~N.~Pyrkov for useful discussion and comments.
This work is partially supported by the program of RAS 
''Element base of quantum computers'',  by the Russian Foundation for Basic Research, grants No.15-07-07928 and 
No.13-03-00017 (A.I.Z.).


\begin{thebibliography}{99}
 
\bibitem{Bose}
S. Bose, Phys. Rev. Lett. {\bf   91}, 207901 (2003)

\bibitem{CDEL}
 M.Christandl, N.Datta, A.Ekert and A.J.Landahl, Phys.Rev.Lett. {\bf   92}, 187902 (2004)

\bibitem{ACDE}
 C.Albanese, M.Christandl, N.Datta and A.Ekert, Phys.Rev.Lett. {\bf   93}, 230502 (2004)

\bibitem{KS}
 P.Karbach and J.Stolze, Phys.Rev.A {\bf   72}, 030301(R) (2005)

 
\bibitem{GKMT}
 G.Gualdi, V.Kostak, I.Marzoli and P.Tombesi, Phys.Rev. A {\bf   78}, 022325 (2008)

 
 
\bibitem{WLKGGB}
A.W\'ojcik, T.Luczak, P.Kurzy\'nski, A.Grudka, T.Gdala, and M.Bednarska
Phys. Rev. A {\bf   72}, 034303 (2005)
 
 
 
 

\bibitem{NJ}
G.M.Nikolopoulos and I.Jex, eds., {\it Quantum State Transfer and Network Engineering}, Series in
Quantum Science and Technology, Springer, Berlin Heidelberg (2014)


 \bibitem{SAOZ}
 J.Stolze, G. A. \'Alvarez,
O. Osenda, A. Zwick in
{\it Quantum State Transfer and Network Engineering.
Quantum Science and Technology},
ed. by  G.M.Nikolopoulos and I.Jex, Springer Berlin Heidelberg, Berlin, p.149  (2014) 




\bibitem{ZLZDLL}
J.Zhang, G. L. Long,  W. Zhang,  Zh. Deng,  W. Liu, and Zh. Lu, Phys.Rev.A {\bf 72}, 012331 (2005)


\bibitem{CRMF}
G. De Chiara, D. Rossini, S. Montangero, R. Fazio, Phys. Rev. A {\bf   72}, 012323 (2005)

\bibitem{ZASO}
 A. Zwick, G.A. \'Alvarez,
J. Stolze, O. Osenda, Phys. Rev. A {\bf   84}, 022311 (2011)
 
 \bibitem{ZASO2}
 A. Zwick, G.A. \'Alvarez,
J. Stolze, O. Osenda, Phys. Rev. A {\bf   85}, 012318 (2012)

\bibitem{ZASO3}
A. Zwick, G.A. \'Alvarez, J. Stolze, O Osenda,
Quant. Inf. Comput. {\bf 15}, 582 (2015)
 


\bibitem{Wootters}
 W.K. Wootters,,
Phys. Rev. Lett. {\bf   80},
2245 (1998)

\bibitem{HW}
S.Hill and W.K.Wootters, Phys. Rev. Lett. {\bf    78}, 5022 (1997)



\bibitem{P}
A.Peres, Phys. Rev. Lett. {\bf    77}, 1413 (1996)


\bibitem{AFOV}
L.Amico, R.Fazio, A.Osterloh and V.Ventral, Rev. Mod. Phys. {\bf    80}, 517 
(2008)

\bibitem{HHHH}
R.Horodecki,
P.Horodecki, M.Horodecki and K.Horodecki, Rev. Mod. Phys. {\bf    81}, 865  (2009)


\bibitem{DFZ}
S.I.Doronin, E.B.Fel'dman, and A.I.Zenchuk,
Phys. Rev. A {\bf 79}, 042310 (2009) 

 
\bibitem{DZ}
 S.I.Doronin, A.I.Zenchuk, Phys. Rev. A {\bf 81}, 022321 (2010)  	


 \bibitem{BACVV2010}
 L. Banchi, T. J. G. Apollaro,  A. Cuccoli,  R. Vaia, and P. Verrucchi, 
 Phys.Rev.A {\bf 82}, 052321 (2010)
 
 \bibitem{BACVV2011}
 L. Banchi, T. J. G. Apollaro, A. Cuccoli, R. Vaia
and P. Verrucchi,
 New J. Phys. {\bf 13}, 123006 (2011) 

 
\bibitem{LS}
P. Lorenz,  J. Stolze,
Phys. Rev. A {\bf 90}, 044301 (2014)

 
\bibitem{BBVB}
L.Banchi,  A. Bayat,  P. Verrucchi, and S.Bose, 
Phys.Rev.Let. {\bf 106}, 140501 (2011)
 
 

\bibitem{BZ}
 B. Chen, and  Zh. Song,
Sci. China-Phys., Mech. Astron {\bf 53}, 1266 (2010)


 
 
 
 
\bibitem{Banchi}
L. Banchi,
Eur. Phys. J. Plus {\bf 128}, 137 (2013) 

\bibitem{YGQ}
W. Qin, J. L. Li, G. L. Long,
Chin. Phys. B {\bf 24}, 040305 (2015) 

\bibitem{YGQ2}
Zh. Yang, M. Gao, W. Qin,
arXiv:1503.06274



\bibitem{QWZ}
W. Qin,  Ch. Wang,  and X. Zhang, Phys.Rev.A {\bf 91}, 042303 (2015)


\bibitem{Z_2012}
A.I.Zenchuk,
J. Phys. A: Math. Theor. {\bf   45} (2012) 115306

 

\bibitem{Z_2014}
A.I.Zenchuk, 
Phys. Rev. A {\bf 90}, 052302(13) (2014) 


 
\bibitem{KF}
 E.I.Kuznetsova and E.B.Fel'dman, J.Exp.Theor.Phys. {\bf   102}, 882 (2006)

 
\bibitem{KZ_2008}
 E.I.Kuznetsova and A.I.Zenchuk, Phys.Lett.A {\bf   372},  pp.6134-6140 (2008)



\bibitem{FZ}
E.B.Fel'dman and A.I.Zenchuk,
Phys. Lett. A {\bf   373} (2009) 1719
 
\bibitem{ZZHE}
M.Zukowski, A.Zeilinger, M.A.Horne, A.K.Ekert, Phys. Rev.
Lett. {\bf 71}, 4287 (1993)

 
\bibitem{BPMEWZ}
D.Bouwmeester, J.-W. Pan, K.Mattle, M.Eibl, H.Weinfurter, and  A. Zeilinger, 
Nature {\bf 390}, 575 (1997)

\bibitem{BBMHP}
D. Boschi,  S. Branca,  F. De Martini, L. Hardy,  and S. Popescu,
Phys. Rev. Lett. {\bf 80}, 1121 (1998)

 
\bibitem{PBGWK2}
N.A.Peters, J.T.Barreiro,  M.E.Goggin, T.-C.Wei,  and P.G.Kwiat, Phys.Rev.Lett. {\bf   94}, 
150502 (2005) 

\bibitem{PBGWK}
N.A.Peters, J.T.Barreiro, M.E.Goggin, T.-C.Wei, and P.G.Kwiat in {\it Quantum
Communications and Quantum Imaging III}, ed. R.E.Meyers,
Ya.Shih, Proc. of SPIE {\bf   5893} (SPIE, Bellingham, WA, 2005) 

\bibitem{DLMRKBPVZBW}
B.Dakic, Ya.O.Lipp, X.Ma, M.Ringbauer, S.Kropatschek,
S.Barz, T.Paterek, V.Vedral, A.Zeilinger, C.Brukner, and P.Walther, 
Nat. Phys. {\bf   8}, 666 (2012). 

\bibitem{XLYG}
G.Y. Xiang, J.Li, B.Yu, and G.C.Guo
Phys. Rev. A {\bf   72}, 012315  (2005)

 
\bibitem{BBCJPW}
C.H.Bennett, G.Brassard, C.Cr\'epeau, R.Jozsa, A.Peres, and W.K.Wootters,
Phys. Rev. Lett. {\bf   70}, 1895 (1993)

 
 
 
 
\bibitem{YS1}
B.Yurke, D.Stoler, Phys. Rev. A {\bf 46}, 2229 (1992)

\bibitem{YS2}
B.Yurke, D.Stoler, Phys. Rev. Lett. {\bf 68}, 1251 (1992)

 
 

\bibitem{BDSSBW}
C.H.Bennett, D.P.DiVincenzo, P.W.Shor, J.A.Smolin, B.M.Terhal, and W.K.Wootters,
Phys.Rev.Lett. {\bf   87}, 077902 (2001);
 Erratum,
 C.H.Bennett, D.P.DiVincenzo, P.W.Shor, J.A.Smolin, B.M.Terhal, and W.K.Wootters, 
 Phys. Rev. Lett. {bf   88}, 099902(E) (2002)
 
 
\bibitem{BHLSW}
C.H.Bennett, P.Hayden,
D.W.Leung, P.W.Shor, and A.Winter, 
IEEE Transetction on Information Theory {\bf   51}, 56 (2005)  

 
 
\bibitem{G}
G.L.Giorgi, Phys. Rev. A {\bf   88}, 022315 (2013) 



\bibitem{Z0}
W. H. Zurek, Ann. Phys.(Leipzig), {\bf 9}, 855 (2000)
 
\bibitem{HV}
L.Henderson and V.Vedral J.Phys.A:Math.Gen. {\bf 34}, 6899 (2001)


\bibitem{OZ}
H.Ollivier and W.H.Zurek, Phys.Rev.Lett. {\bf 88}, 017901 (2001) 

\bibitem{Z}
W. H. Zurek, Rev. Mod. Phys. {\bf 75}, 715 (2003)





\bibitem{DSC}
Datta, A., Shaji, A., Caves, C.M.,
Phys. Rev. Lett. {\bf 100}, 050502 (2008)

\bibitem{LBAW}
 Lanyon, B.P., Barbieri, M., Almeida, M.P., White, A.G.,
 Phys. Rev. Lett. {\bf 101}, 200501 (2008)

\bibitem{NLLZ}
W.J.Nie, Yu.H.Lan, Yo.Li, and Sh.Ya.Zhu, 
Sci.China-Phys., Mech. Astron {\bf 57}, 2276 (2014)

\bibitem{ZC}
P. Zhang, B. You, and L.-X. Cen,
Chin. Sci. Bull., {\bf 59}, 3841 (2014)

\bibitem{SXSZDWHCKW}
J.X. Sci, W. Xu, G. Sun et al,  Chin. Sci. Bull. {\bf 59}, 2547 (2014)

 \bibitem{RDL}
 S. Rodriques, N. Datta, and P. J. Love,
 Phys. Rev. A {\bf 90}, 012340 (2014) 




\end{thebibliography}
\end{document}